\documentclass[journal=jacsat,manuscript=article]{achemso}

\usepackage{chemformula} 
\usepackage[T1]{fontenc} 
\usepackage{graphics}      
\usepackage{mciteplus}
\usepackage{natbib}
\usepackage{xcolor}
\usepackage{soul}

\usepackage{wrapfig}
\usepackage{lipsum}
\usepackage{float}
\usepackage{amsmath}
\usepackage{hyperref}

\usepackage{amsmath,amsfonts,amssymb}

\newcommand{\figref}[1]{Fig.~\ref{fig:#1}}

\DeclareMathOperator*{\argmax}{arg\,max}

\newcommand{\figlight}{
\begin{figure*}[t]
\centering
\includegraphics[trim={0in 0in 0in 0},width=\linewidth]{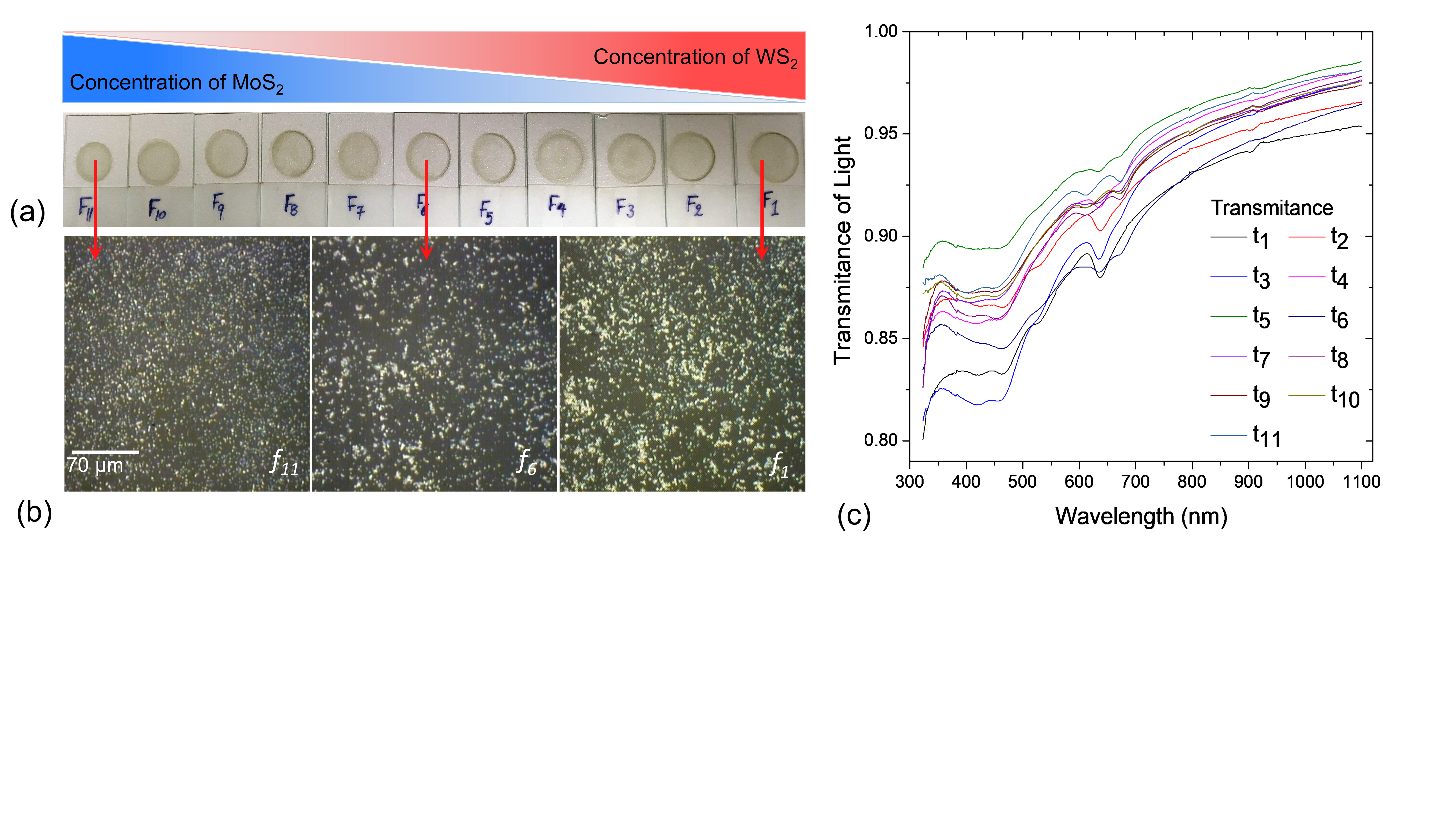}

\vspace{-1.3in}
\caption{ (a) Eleven filters drop-casted on glass slides; $f_{1}$ is 100\% $\text{WS}_{2}$, but $f_{2},\dots,f_{10}$ are made by gradually adding $\text{MoS}_{2}$ and decreasing $\text{WS}_{2}$, and finally $f_{11}$ is 100\% $\text{MoS}_{2}$. (b) Microscopic image of three filters $f_{1}$, $f_{6}$, $f_{11}$: nanomaterials on glass substrate. (c) Background-subtracted transmittance $vs.$ wavelength $t_{1}$, ..., $t_{11}$ for all 11 filters. The excitonic peaks get modified gradually from $f_{1}$ to $f_{11}$ as a result of changing proportion of mixing two TMDs. (The figure is re-plotted from the original Hejazi \emph{et al.,} 2019\cite{hejazi2019transition}).}
\label{fig:light}
\end{figure*}
}

\newcommand{\figmethods}{
\begin{figure*}
\centering
\includegraphics[trim={0in 0in 0in 0},width=1.1\linewidth]{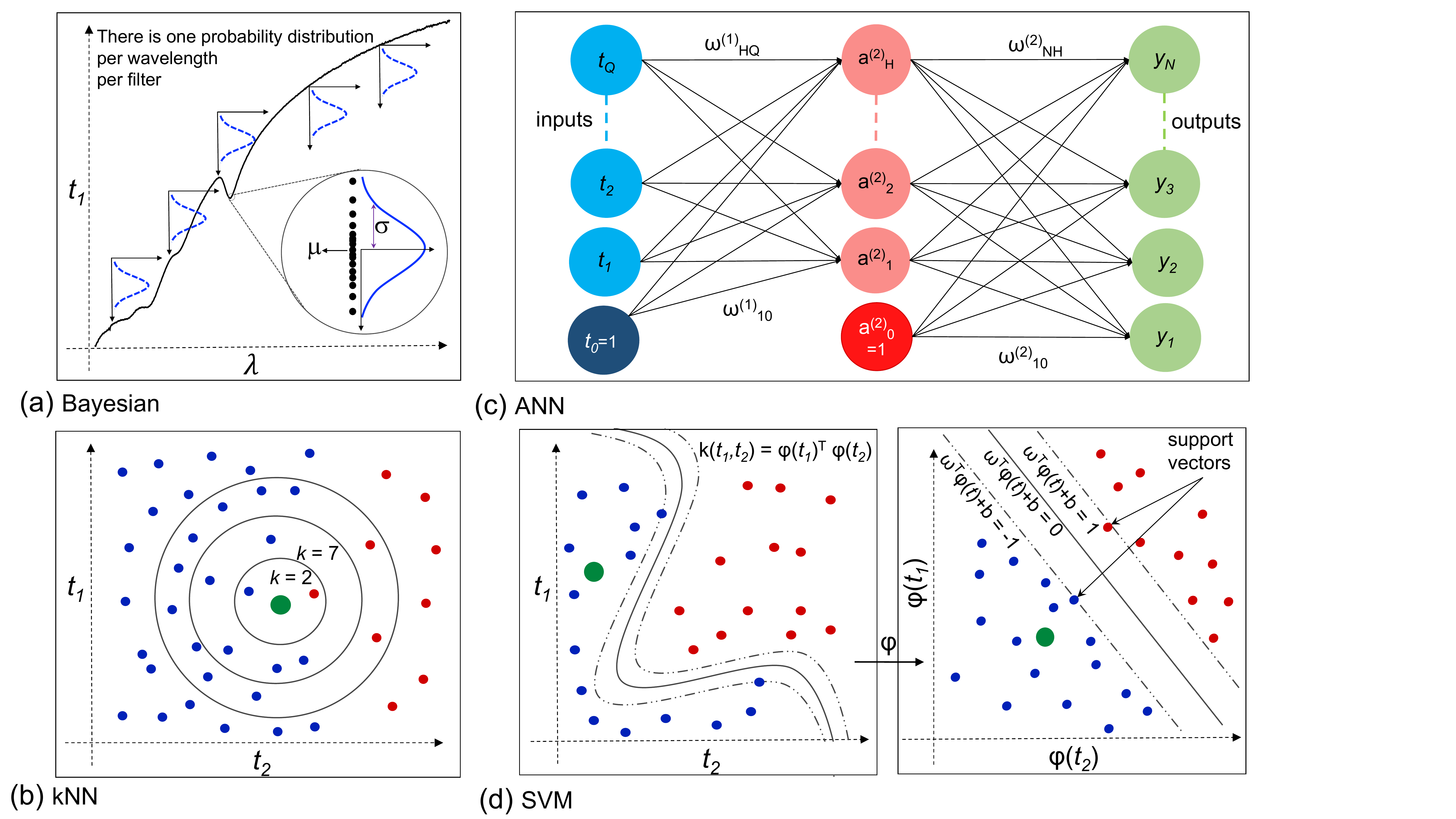}
\vspace{-0.2in}
\caption{Schematic representations that distinguish the various analysis approaches used in this work: (a) The Bayesian inference, that shows at each wavelength over each filter a probability distribution can be formed from the labeled samples of that given wavelength. (b) The kNN algorithm. Each point represents a sample in an 11-dimensional space (transmittances $t_1,...,t_{11}$), but only two dimensions are shown for convenience. Blue and red points represent samples belonging to two  wavelengths that have close transmittance values. The unknown sample in Green will be classified depending on the majority votes of the samples encircled in the circles depending on the number of the closest neighbors. With $k=2$ the choice is not certain but with $k=7$ the unknown sample obviously belongs to the class of Blue points. (c) The fully-connected three layered ANN model. Each neuron is connected to the neurons in previous layer via weight parameters that must be optimized for the model to correctly estimate the unknown samples class. The bias neurons (in dark Blue and dark Red) are not connected to previous layers since they are by definition equal to $+1$. (d) The non-linear SVM algorithm, where the wavelength classes are the same as (b) and the Gray solid line draws the barrier between the two classes. The dashed lines indicate the margins. By doing a kernel trick we can transform the data from feature space $t$ to its dual space $\phi(t)$.}
\label{fig:methods}
\end{figure*}
}

\newcommand{\figAllerrors}{
\begin{figure}[t]
\centering
\vspace{-0.5in}
\includegraphics[trim={0in 0in 0in 0},width=\linewidth]{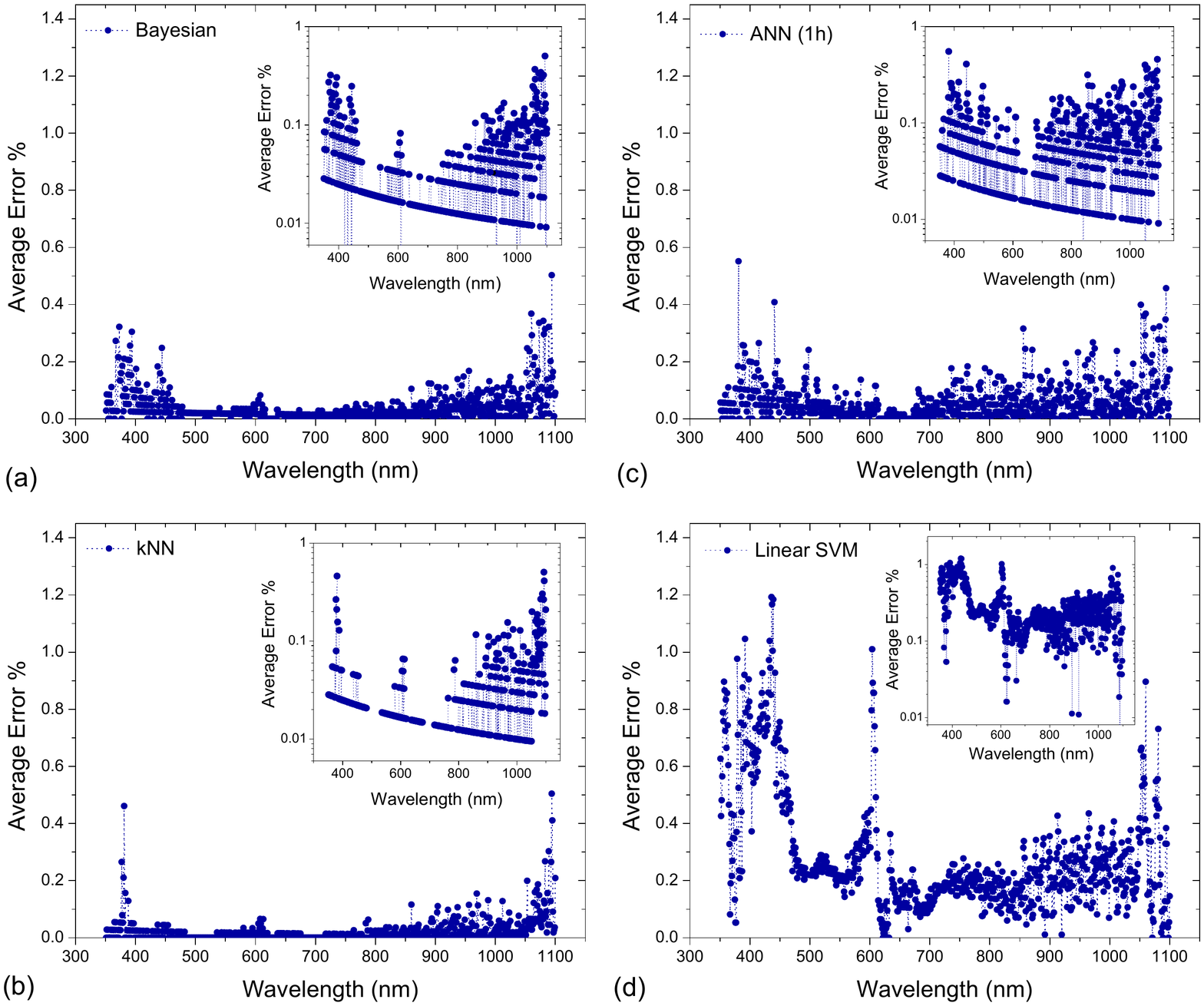}

\caption{The percent error of estimating wavelength of test samples in a given wavelength averaged over 10 samples of the same wavelength; we define the estimation error percent as $error\% = \frac{\mid \lambda_{Groundtruth}(nm)-\lambda_{Estimated}(nm)\mid}{\lambda_{Groundtruth}(nm)}\times{100}$. (a) Bayesian inference; (b) kNN algorithm; (c) ANN algorithm; and (d) Linear SVM, using all 11 filters. The insets are semi-log plot of the same figures. The y-axis of inset plots have been limited by cutting of the values that are too close to 0, for better visibility, and consequently these points do not show up in semi-log plot.}
\label{fig:Allerrors}
\end{figure}
}

\newcommand{\figAbserrorTesttime}{
\begin{figure*}[t]
\centering
\includegraphics[trim={0in 0in 0in 0},width=\linewidth]{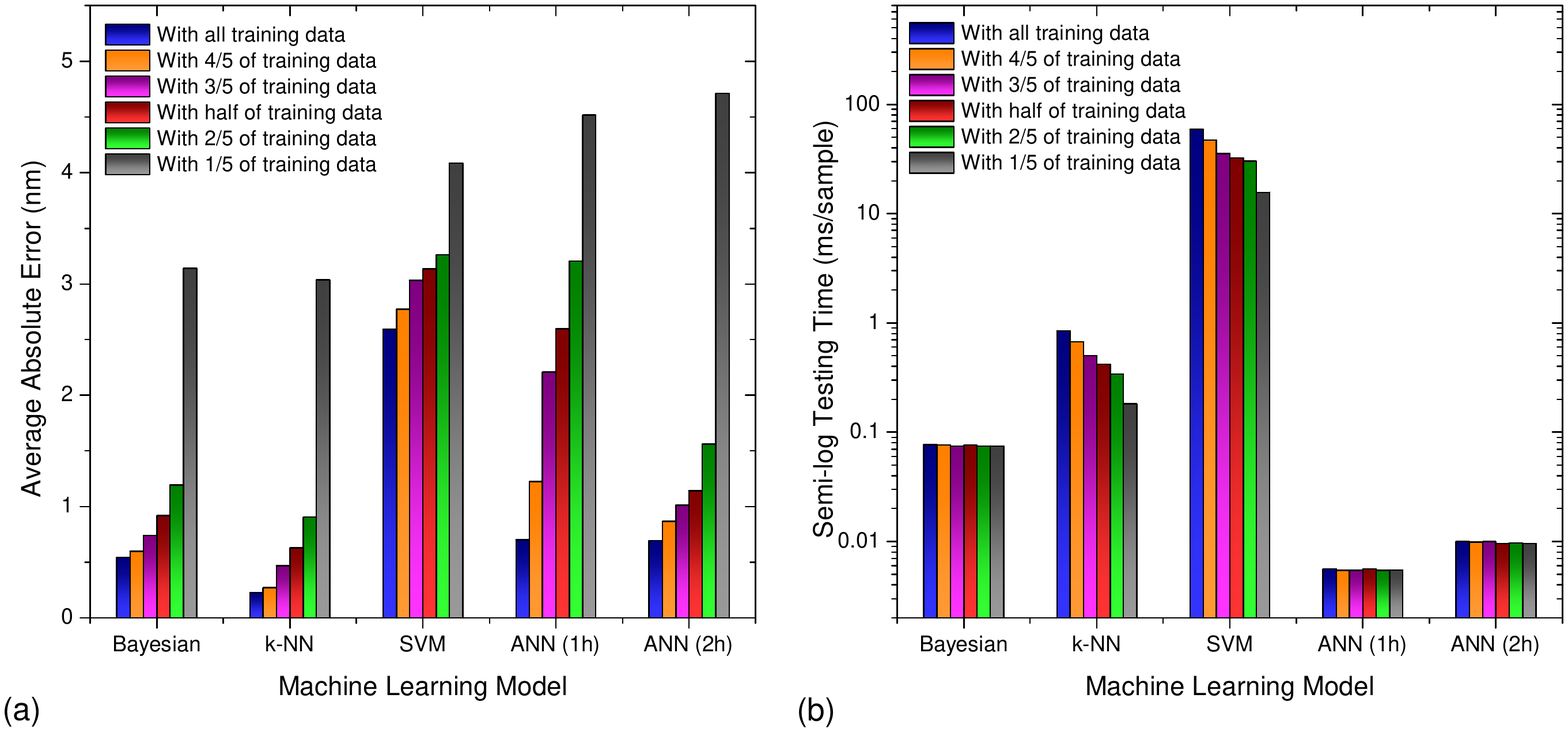}
\vspace{-2.1in}
\caption{(a) Average absolute error calculated by averaging $abs.err=|\lambda_{Real}(nm)-\lambda_{Estimated}(nm)|$ of all 7500 test samples when different sizes of randomly chosen training data are used for training the model, performed with each of the 4 machine learning methods; (b) The semi-log plot of required time to test each new sample using the trained models after all training steps are completed.}
\label{fig:AbserrorTesttime}
\end{figure*}
}

\newcommand{\figOldnewDrift}{
\begin{figure*}[t]
\centering
\includegraphics[trim={0in 0in 0in 0},width=\linewidth]{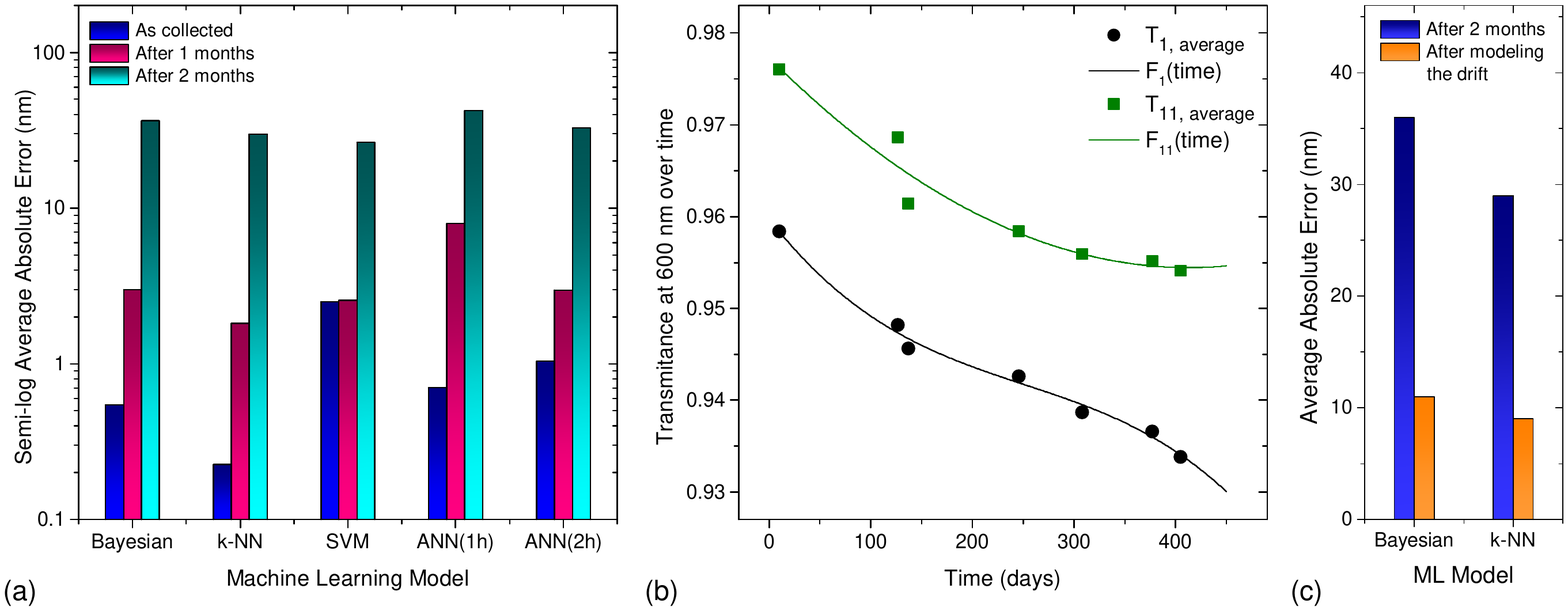}

\vspace{-2.4in}
\caption{(a) Average absolute error of estimation in semi-log scale using all training data for test samples collected at the same time as the training data, compared to the test samples collected after one and two months; (b) Third order polynomial functions fitted to the average transmittance of filters $f_1$ and $f_{11}$ over period of $\sim 400$ days. Scatter plots are the average measured transmittance values and the solid lines indicated the fitted functions; (c) Average absolute error of estimating wavelength of test samples collected two months after training when no modification is applied to the model (Blue bars), and when the training sample-label pairs are corrected using the drift functions (Orange bars) in Bayesian and kNN models.}
\label{fig:OldnewDrift}
\end{figure*}
}

\newcommand{\figTOC}{
\begin{wrapfigure}{r}{0.46\textwidth}
\vspace{-.2in}
\centering
\includegraphics[trim={0in 1.2in 1in 0},width=\linewidth]{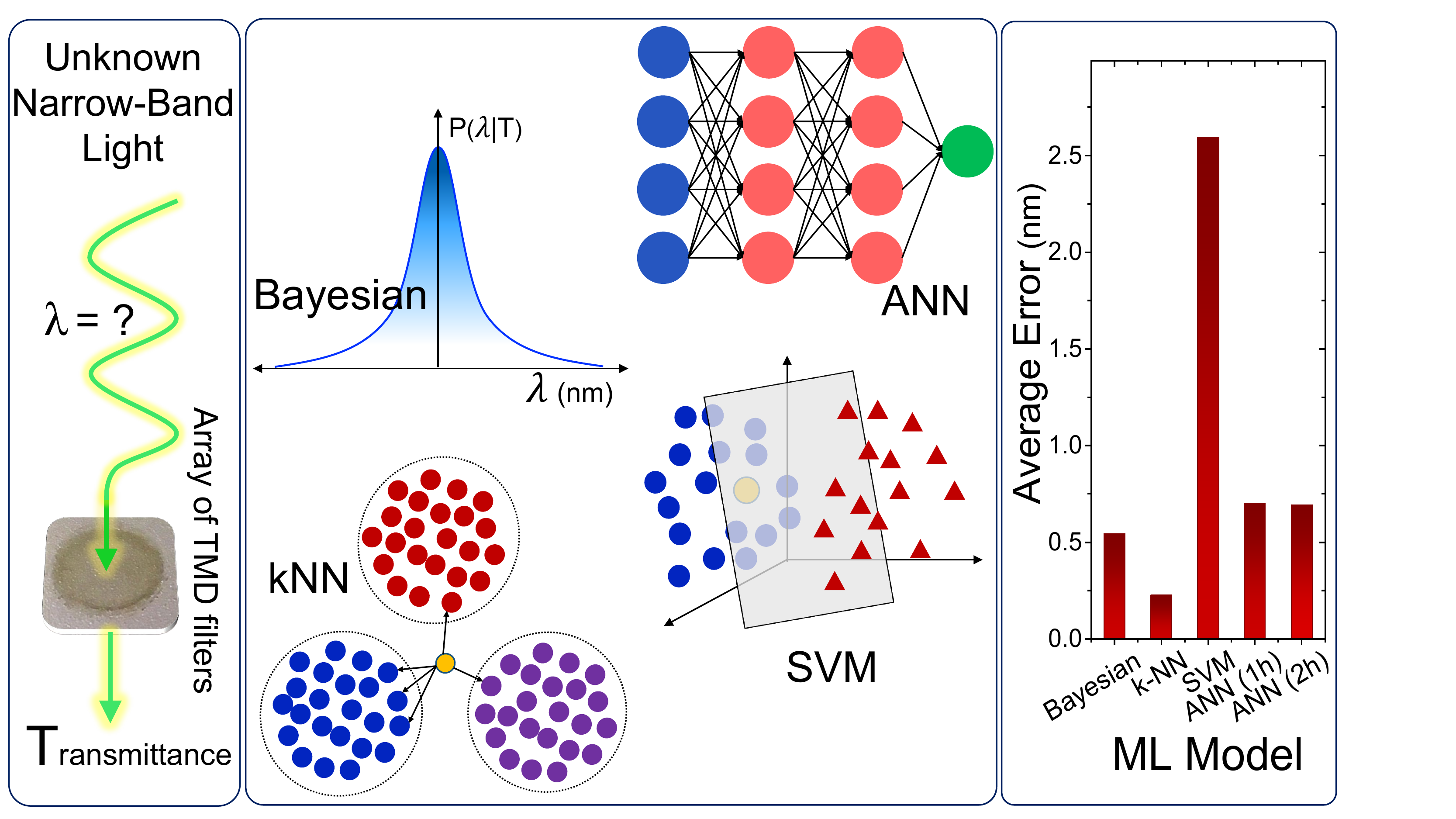}
\end{wrapfigure}
}

\author{Davoud~Hejazi}
\affiliation{Department of Physics, Northeastern University, Boston, MA, USA}

\author{Shuangjun~Liu}
\affiliation{Augmented Cognition Lab (ACLab), Electrical and Computer Engineering Department, Northeastern University, Boston, MA, USA}

\author{Amirreza~Farnoosh}
\affiliation{Augmented Cognition Lab (ACLab), Electrical and Computer Engineering Department, Northeastern University, Boston, MA, USA}

\author{Sarah~Ostadabbas}
\affiliation{Augmented Cognition Lab (ACLab), Electrical and Computer Engineering Department, Northeastern University, Boston, MA, USA}
\email{*ostadabbas@ece.neu.edu}
\author{Swastik~Kar}
\affiliation{Department of Physics, Northeastern University, Boston, MA, USA}
\email{*s.kar@northeastern.edu}

\title{Development of Use-specific High Performance Cyber-Nanomaterial Optical Detectors by Effective Choice of Machine Learning Algorithms}

\begin{document}



\section{Abstract}
\figTOC
Due to their inherent variabilities, nanomaterials-based sensors are challenging to translate into real-world applications, where reliability and reproducibility is the key. Recently we showed that Bayesian inference can be employed on engineered variability in layered nanomaterials-based optical transmission filters to determine optical wavelengths with ultra-high accuracy and precision. In many practical applications, however, the sensing cost/speed and long-term reliability can be equal or more important considerations. Although various machine learning (ML) tools are frequently used on sensor and detector networks to address these considerations and dramatically enhance their functionalities, nonetheless, their effectiveness on nanomaterials-based sensors has not been explored.  Here, we show that the best choice of ML algorithm in a cyber-nanomaterial detector is largely determined by the specific use-considerations, including accuracy, computational cost, speed, and resilience against drifts and long-term ageing effects. When sufficient data and computing resources are provided, the highest sensing accuracy can be achieved by the k-nearest neighbors (kNN) and Bayesian inference algorithms, however, these algorithms can be computationally expensive for real-time applications. In contrast, artificial neural networks (ANN) are computationally expensive to train (off-line), but they provide the fastest result under testing conditions (on-line) while remaining reasonably accurate. When access to data is limited, support vector machines (SVMs) can perform well even with small training sample sizes, while other algorithms show considerable reduction in accuracy if data is scarce, hence, setting a lower limit on the size of required training data. We also show by tracking and modeling the long-term drifts of the detector performance over large (i.e. one year) time-frame, it is possible to dramatically improve the predictive accuracy without the need for any recalibration. Our research shows for the first time that if the ML algorithm is chosen specific to the use-case, low-cost solution-processed cyber-nanomaterial detectors can be practically implemented under diverse operational requirements, despite their inherent variabilities. 


\vspace{0.3in}
\textbf{Keywords:} 2D materials, artificial neural networks (ANN), Bayesian inference, k-nearest neighbor (kNN), layered materials, machine learning, optical detectors, optical wavelength estimation, semiconductors, support vector machine (SVM), transition metal dichalcogenides (TMDs)

\section{Introduction}
Nanomaterials are very attractive for building sensors, and various examples of using 2D nanomaterials, nano-tubes, quantum-dots, $etc.$, can be found in the fabrication of optical detectors\cite{west2003engineered, rao2019enhanced, wang2019recent}, molecular and bio-sensors\cite{yang2010carbon, wang2005nanomaterial, mojtabavi2019single, liu2019peroxidase, alhamoud2019advances}, ion and radiation sensors\cite{hao2019ion},chemical sensors\cite{galstyan2019highly, meng2019electrically} gas sensors\cite{li2005co3o4, hennighausen2019oxygen}, temperature sensors\cite{yadav2008moisture} and many other cases of detection and sensing. There are many aspects that make nanomaterials promising candidates for these applications compared to the bulk materials. For instance, their enhanced optoelectronic and novel chemical/physical properties make them efficient choices for sensing, while their small dimensions will lead to devices with lower power consumption and smaller size. In many cases, nanomaterials are much more attractive than conventional semiconductor sensors due to their low-cost, earth-abundant availability, and compatibility with affordable solution-processable techniques. Their high surface-to-volume ratio makes them highly sensitive as chemical sensors, whereas their quantum confinement or excitonic processes enables them to be excellent target-specific photodetectors. As a result, over the past decades, there has been a tremendous progress in fundamental understanding and proof-of-concept demonstrations of chemical, biological, optical, radiological and a variety of other sensors using nanomaterials\cite{west2003engineered, rao2019enhanced, wang2019recent, yang2010carbon, wang2005nanomaterial, liu2019peroxidase, alhamoud2019advances, hejazi2019wavelength, hejazi2019bayesian, galstyan2019highly, meng2019electrically, li2005co3o4, hennighausen2019oxygen, yadav2008moisture, hao2019ion}.

However, there exists many challenges in real-world implementation of sensors made from nanomaterials; above all the difficulties in reproducing them which makes the size and physical location of the fabricated nanomaterials on the substrate unpredictable and uncontrolable. Moreover, the nanomaterials undergo gradual decay in ambient condition called "drift", $i.e.$ they are not very stable; also there is often a large noise in their measurement because of their small size due to the fact that nanomaterials not only respond to what they are designed to measure, but also are very sensitive to many other conditions in their environment. These shortcomings, not to mention the gradual decays of nanomaterials, have introduced huge challenges in mass production of reliable devices from them, where predictable and controllable manufacturing processes is essential to the industry.

In recent decades, the emergence of machine learning (ML) has demonstrated a great potential for enhancing statistical analysis in the field of material science. Nowadays, ML provides popular tools for obtaining information from internet of things (IoT) networks \cite{lee2015internet, sharma2019towards, lynggaard2019controlling, hussain2019machine, yang2019machine} such as charge-coupled devices (CCDs)\cite{carbune2019predicted, nai2019real}, complementary metal-oxide-semiconductor (CMOS) detectors\cite{yan2019novel, guan2019high, brunckhorst2019machine}, or regular Silicon-based spectrometers, which are examples of sophisticated networks of optical detectors\cite{butler2018machine, schmidt2019recent}. In physics, on one hand, people employ machine learning to analyze, predict, or interpret physical quantities; on the other hand,  underlying physical principle has also been employed to facilitate designing effective machine learning tools\cite{liu2019seeing, farnooshintroduction}. ML methods have been successfully applied for accelerated discovery\cite{xue2016accelerated, yuan2018accelerated, raccuglia2016machine} and development of materials and metamaterials with targeted properties\cite{ley2015organic, correa2018accelerating, iwasaki2019materials, malkiel2018plasmonic, mlinar2013engineered, ma2018deep, bacigalupo2016design}, predicting chemical\cite{coley2017prediction, padula2019combining, hoyt2019machine, schlexer2019machine} and optoelectronic properties of materials\cite{goldberg2015prediction, ma2019accelerated, huang2019band, khabushev2019machine}, and synthesizing nanomaterials\cite{tang2019machine}. The variations in nanomaterial properties are usually considered as "noise" and various experimental or statistical approaches are often pursued to reduce these variations or to capture the useful target data from noisy measurements\cite{mihaila20041, ru2010automated, sikula2006advanced, wu2008high, guiot2009measurement}. However, the direct applications of the data analytic approaches have never been sought on the variability of the nanomaterials themselves to utilize these variations as information instead of treating them as noise. In the context of sensing applications, one way to overcome the aforementioned challenges of nanomaterials is to use ML on a multitude of sensors in order to extract relevant response patterns towards achieving  accurate, reliable, and reproducible sensing outcomes. 

\figlight

Our previous work (Hejazi \emph{et al.,} 2019\cite{hejazi2019transition, hejazi2019wavelength}) demonstrated the power of using advanced data analytic on the measured data from a few uncontrolled low-cost, easy-to-fabricate semiconducting nanomaterials in order to estimate peak wavelength of any incoming monochromatic/near monochromatic light over the spectrum range of 351--1100 $nm$ with high precision and accuracy, in which we created the world's first cyber-physical optical detector. In that work, we applied a Bayesian inference on optical transmittance data of 11 nanomaterial filters fabricated from two transition-metal dichalgogenides, $MoS_2$ and $WS_2$ (see \figref{light}). We were also able to reduce the number of filters to two filters via step-wise elimination of least useful filters and still achieve acceptable results even with two filters. We also discussed that it is possible to choose suitable materials for desired spectrum ranges for optical filter fabrication.

In the present work, our aim is to augment our analytical tools by employing various ML techniques, compare their efficacy in color sensing, and finally choose the most suitable ML algorithm for color detection based on the application requirements. We note in doing so, it is important to discuss the data-analytical process of ML techniques within the context of nanoscience datasets, so that they can be appropriately utilized in analyzing nanoscience data of other types as well. Hence, we provide below a brief outline, using schematic visualizations, of how different ML approaches are analyzing our data. When ML is used as a discriminative model in order to distinguish different categories ($e.g.$ different optical wavelengths), it comes in one of these two forms: "supervised learning", where new samples are classified into $N$ categories through training based on the existing sample-label pairs; and "unsupervised learning", where the labels are not available, and the algorithm tries to cluster samples of similar kind into their respective categories. In our target application in this paper, labels are wavelengths that combined with measured transmittance values that we will call filter readings, create the set of sample-label pairs known as the training set. Therefore, we chose our analytical approaches based on the supervised ML algorithms. Apart from the Bayesian inference, we employed k-nearest neighbour (kNN), artificial neural networks (ANN), and support vector machines (SVM); the details of each can be found in the Computational Details section. In the following discussions, we provide a brief overview of each method to clarify their algorithmic steps. 

\figmethods

As for the Bayesian inference, we discussed its underlying statistical approach in details in our previous article \cite{hejazi2019transition}. For a given set of known sample-label pairs (i.e. training set), Bayesian inference gathers statistics of the data and uses them later to classify an unknown new sample by maximizing the collective probability of the new sample belonging to corresponding category (see \figref{methods}(c)).

In pattern recognition, the kNN is a non-parametric supervised learning algorithm used for classification and regression\cite{altman1992introduction}, which searches through all known cases and classifies unknown new cases based on a similarity measure defined as a norm-based distance function ($e.g.$ Euclidean distance or norm 2 distance). Basically, a new sample is classified into a specific category when in average that category's members have smallest distance from the unknown sample (see \figref{methods}(a)). Here, $k$ is the number closest cases to the unknown sample, and extra computation is needed to determine the best $k$ value. This method can be very time-consuming if the data size ($i.e.$ total number of known sample-label pairs) is large.

ANNs are computing models that are inspired by, but not necessarily identical to, the biological neural networks. Such models "learn" to perform tasks by considering samples, generally without being programmed with any task-specific rules. An ANN is based on a collection of connected units or nodes called artificial neurons, that upon receiving a signal can process it and then pass the processed signal to the additional artificial neurons connected to them. A neural network has always an \emph{input} layer that are the features of each training sample and an \emph{output} layer that are the classes in classification problem, while it can also be only a number in regression problem. However, there are often more than just two layers in an ANN model. The extra layers that are always located between the input and output layers are called \emph{hidden} layers. The number of hidden layers, the number of neurons in each layer, and how these layers are connected form the neural network architecture\cite{hsu1995artificial, kavzoglu1999determining, wilamowski2007neural}. In general, having more number of hidden layers increases the capacity of the network to learn more details from the available dataset, but having much more layers than necessary can result in overfitting the model to the training set $i.e.$ the model might be performing well on the training set but poorly on the unseen test set\cite{montufar2014number, kavzoglu1999determining}. In this work we have used two different fully-connected ANN architectures to investigate their efficacy on optical wavelength estimation. The schematics of a three layered fully-connected ANN model is shown in \figref{methods}(d). Backpropagation is the central mechanism by which a neural network learns. An ANN propagates the signal of the input data forward through its parameters called weights towards the moment of decision, and then backpropagates the information about error, in reverse through the network, so that it can alter the parameters. In order to train an ANN and find its parameters using the training set, we give labels to the output layer, and then use backpropagation to correct any mistakes which have been made until the training error becomes in an acceptable range \cite{bishop2006pattern}.

When it comes to supervised classification, SVM algorithms are among the powerful ML inference models\cite{cortes1995support, bishop2006pattern, mangasarian2001lagrangian}. In its primary format as a non-probabilistic binary linear classifier, given labeled training data, SVM outputs an optimal hyperplane which categorizes new examples into two classes. This hyperplane is learned based on the "maximum margin" concept in which it divides the (training) examples of separate categories by a clear gap that is as wide as possible \cite{ng2017machine, osuna1997improved}. When we have more than two classes, SVM can be used as a combination of several one vs. rest classifiers to find hyperplanes that discriminate one category from the rest of them. SVM can also efficiently perform non-linear classification using what is called the kernel method by implicitly mapping the samples original features set into a higher dimensional feature space, as illustrated in \figref{methods}(b), and a new sample is classified depending on the side of the margin that it falls in. In this paper, apart from linear SVM, five choices of kernels are examined when using SVM classifiers.

In real-world sensing and other "estimation" applications, the needs ($i.e.$ speed, accuracy, low-complexity etc.) of the end-use should determine the approach or method. Keeping these in mind, we have compared the efficacy of these ML technique by considering the following main considerations: (a) The average error in estimating wavelength of test samples collected at the same time the training samples were collected; (b) The average absolute error for entire spectrum; (c) The required time for training; (d) The elapsed time for estimating wavelength of one test sample using model/trained parameters; (e) The effect of reducing the training set size on efficacy of each model; and (f) How well the models behave on new set of test samples collected several months after the training. Applying these four ML techniques to our wavelength estimation problem has revealed important facts about their efficacy. The kNN algorithm appears to perform the best in terms of the estimation accuracy, however unlike the other three techniques, kNN time complexity is directly proportional to the size of the training set, which will hinder its use in applications that demand real-time implementation. It is due to the fact that kNN is a non-parametric algorithm, in which the model parameters actually grows with the training set size. $k$ should be considered as hyper-parameter in kNN. On the other hand, ANN models perform fastest in the test time, since all of the model parameters in ANNs are learned from the data during the training time, and the test time is only the classification step, which is simply calculating the output value of an already-learned function. Typically, larger training set improves ANN's performance since it leads to a model that is more generalizable to an unseen test data. An interesting observation from our results is that the SVM model shows slightly larger estimation errors compared to the rest of the algorithms, however it is not sensitive to data size and is more resistant to time-dependent variations in optoelectronic response of nanomaterials $i.e.$ to drift. Bayesian inference turns out to be very accurate, and quite fast as well.

By looking at the outcomes of our estimation problem, we have also discovered another important aspect of the data that we are dealing with in the nanomaterial applications. We noticed a significant nanomaterial measurements drift over time in our dataset, which can be described as "evolving class distributions". This means the same object ($i.e.$ light ray) will not create the same responses on the nanomaterial filters over time. Therefore, a model trained on a training set may have completely different parameter values compared to the same model trained on another training set collected after a period of time ($e.g.$ a couple of months in our case). In attempt to overcome the shifts in the data due to the drift in electronic and spectral transmittance of nanomaterials, we show that it is possible to model the drift of nanomaterial responses over time and combine them with the future estimations where the nanomaterial filters have drifted even more. By observing the transmittance of filters over the period of more than a year, we were able to predict the drift in transmittance after two months and improve the performance in the wavelength estimation. This was however only possible in the kNN and Bayesian algorithms since they employ no other parameters than the transmittance values themselves, while SVM and ANN train their own corresponding parameters. In the next section we will summarize the main findings of our work.

\section{Results and Discussion}
\label{sec:results}
The detailed description of each ML algorithm, the number of parameters to be trained, and the computational complexity of each technique will be discussed in the Computational Details section. The resolution of the collected wavelength samples is 1 $nm$. To discuss the efficacy of our wavelength estimators, we define the estimation error percent as $error\% = \frac{\mid \lambda_{Groundtruth}(nm)-\lambda_{Estimated}(nm)\mid}{\lambda_{Groundtruth}(nm)}\times{100}$.

We first present our results comparing the wavelength estimation accuracy from various techniques. \figref{Allerrors} shows a comparison of wavelength estimation by different ML techniques performed using the same set of training data comprised of 75,000 samples. The average errors are the average of error percentages for 10 test samples for each wavelength. By comparing the overall values of the average error as a function of wavelength, it is possible to see that the kNN method appears to best estimate the wavelength of an unknown light source, followed by the Bayesian inference method, when the estimation conditions (time, number of filters used, training size etc.) are not constrained to low values. The ANN and SVM are in the $3^{rd}$ and $4^{th}$ place in overall performance on estimating wavelength of test samples. In order to perform a more quantifiable comparison between the various approaches, we have calculated the average absolute error of entire spectrum by calculating the absolute error ($\mid\lambda_{Groundtruth}(nm)-\lambda_{Estimated}(nm)\mid$) for all 7,500 test samples and averaging them (see \figref{AbserrorTesttime}(a)). In addition, we have performed the AAN using both 1 and 2 hidden layers, which has been presented in the comparison data shown in \figref{AbserrorTesttime} and subsequent figures, where we can see a fifth batch of columns for 2 hidden layer ANN shown with AAN(2h) as opposed to ANN(1h) with 1 hidden layer. 

\figAllerrors

To investigate the sensitivity of the models to the size of the training set, we randomly picked different portions of the training set to perform the training and testing, $i.e.$ by randomly choosing $\frac{1}{5}$, $\frac{2}{5}$, $etc.$ of the original dataset (see \figref{AbserrorTesttime}(a)). As it was expected from theory, the SVM model is least sensitive to the size of training data, followed by the Bayesian inference. However, the ANN and kNN show considerable reduction in performance by reducing the training set size. We can see that \figref{AbserrorTesttime}(a) where SVM shows minor changes from one data size to other, while for instance  1 hidden layer ANN shows steep change in error values. Another important fact that we learn from this figure is the minimum size of training set required to perform reasonable estimation. As we see, in each case using only $\frac{1}{5}$ of the training set, the average absolute error tends to increase considerably. The $\frac{1}{5}th$ translates to 20 times the number of different classes (wavelengths in our case), which sets a lower bound on size of the training dataset that must be collected. Another non-trivial and highly interesting observation is the relative errors of 1 hidden layer $vs.$ 2 hidden layer ANN model as the error in wavelength estimation rises more sharply with decreasing training sets in the 1 hidden layer ANN, suggesting that ANN with more hidden layers appears to "learn" better from the available data and yield more accurate estimations. The other consideration is the available data is not exactly enough for this problem even when all data is used. This can be justified by seeing that even from going from $\frac{4}{5}$ to all of the data there is a noticeable change in overall accuracy, while we expect to see minor change in accuracy of each model if the supplied data was sufficient.


\figAbserrorTesttime

We next analyze the performance of each algorithm in terms of the required time for each model to train, and afterwards to test. In kNN and Bayesian models there are no real learning steps, and as a result there is a definitive answer for value of a test sample with a given training set. The kNN model calculates the distance of the test sample from every training samples, which are fixed; so the testing time is directly related to the size of the training set. Given our relatively small dataset the kNN model works rather fast, but most likely it would not be the case if larger dataset were used (see \figref{AbserrorTesttime}(b)). As for Bayesian algorithm, the training part is limited to collecting the statistics from training data. In testing step, the model searches through all probability distributions and maximizes the $posteriori$; though it is obviously time consuming but is independent from the training set size. Hence, in both models the main and/or only required time is for testing.

As for ANN and SVM the training step can be dynamically decided by desired conditions. In the case of SVM, the training step is governed by choice of tolerance, kernel type, $etc.$. After the support vectors are found, the testing step is carried out by checking which side of the hyperplanes the test sample falls. In our study different choices of kernel/tolerance did not pose meaningful enhancement on the estimation efficacy of the trained SVM models.

The situation is quite different for ANN, since one can iterate the training loop infinite times and the results may either improve, converge, or just get stuck in a local minima. Time and computational resources for training are the real costs of the ANN algorithm, but in general ANN can fit very complicated non-linear functions that other models might not have as good performance as ANN. After the end of training step (decided by the experimenter based on the desired level of accuracy), the testing step is basically a few matrix multiplications only, as explained in Computational Details section. Hence, the testing time of ANN is quite short and independent from the size of training set. IN addition, we found that with smaller training sets the ANN model is prone to over-fitting, $i.e.$ the model might perform well on the training set itself but not on new test set. The required testing time for each sample when all training steps are completed is shown in \figref{AbserrorTesttime}(b), which is the more relevant time-scale for real-world applications. The details of each model and their computational complexity is discussed in Computational Details section.

Owing to their affinity for adsorbing oxygen, and moisture, as well as through creation of defects with exposure to ambient conditions, the electronic and optical properties of the nanomaterials gradually evolve with time, which are reflected in drifts in their spectral transmittance values. Hence, even though they show fair stability in short period of time, the effective transmittance of our nanomaterial filters slowly change over time. This causes slow reduction of accuracy in estimating wavelength over time in later measurements. However, as shown in our previous work \cite{hejazi2019transition} by calibrating the filters from time to time it would be possible to continue using these same filters over extended periods of time, and the efficacy of estimations does not suffer from wears or minor scratches, since the re-calibration will overcome the gradual changes of the filters. 

In the current work we instead investigate the performance of each ML method over time by testing the efficacy of the trained models on the new test samples collected after two months. The average absolute error (with all training data) of estimating wavelength of newly collected test samples and original test samples are shown in \figref{OldnewDrift}(a). Quite interestingly, the SVM model shows minimal change in the estimation accuracy having the ratio of $\sim 1$ in first month, and smaller change later, while all other models show considerable reduction in accuracy. This change is quite obvious in a 1 hidden layered ANN. Next, we discuss how to overcome the effect of transmittance change of filters as a result of drift in optoelectronic response of nanomaterials by modeling the drift over time.

\figOldnewDrift

Choosing a proper ML technique that performs more robust over time is only one way of using the filters over time without the need for re-calibration; but it is also possible to model the drift of nanomaterial. For this purpose we observed the transmittance change over time for our nanomaterial filters in a period of about 400 days, and tried to fit a polynomial curve to the average transmittance values at each wavelength for each filter with respect to number of days after the filters were fabricated. Two examples of these curves shown in \figref{OldnewDrift}(b) are for filters $f_1$ and $f_{11}$ at 500 nm, that present the slow decrease of transmittance over time, where a third order polynomial function fairly fits the drift. To check the validity of our claim we calculated the expected transmittance at each wavelength for each filter around the day 450, which was the day that another new set of test samples were collected. In case of Bayesian, we replaced the mean value of transmittance by the calculated transmittance values at the day 450, while for kNN, we multiplied each transmittance $t$ in training set by a corresponding coefficient $\frac{t_{avg}(450)}{t_{avg}(0)}\times t$, then used them for estimation (see \figref{OldnewDrift}(c)). Applying the drift over time functions is only possible for kNN and Bayesian algorithms since they don't have a distinct learning step, while the two other models have already trained their parameters based on the old training data. The results show that Bayesian model is more compatible with the drift over time function, which is expected since these fitting functions are calculated using the mean transmittance values, as Bayesian algorithm also uses mean values/standard deviations for estimation.

\section{Conclusions} 
\label{sec:conclusion}
In conclusion, we have successfully demonstrated the efficacy of various ML techniques in estimating the wavelength of any narrow-band incident light in spectrum range 351--1100 $nm$ with high accuracy using the optical transmittance information collected from a few low-cost nanomaterial filters that require minimal control in fabrication. With the available data the kNN algorithm shows highest accuracy with the average estimation errors reaching to 0.2 $nm$ over the entire 351--1100 $nm$ spectrum range, where the training set is collected with 1 $nm$ spectral resolution; but this method is not suitable for real-time applications since the required testing time is linearly proportional to the training set size. The situation is almost the same with the Bayesian algorithm which performs very well, but although its speed is not data size dependent, still the process is much slower than the other methods. The real-time speed considerations can be very well satisfied with ANN models where the estimation time can be as low as 10$\mu s$, but these models as well as Bayesian and kNN turn out to be more sensitive to drift in spectral transmittance of nanomaterials over time. On the other hand SVM models show a bit lower accuracy compared to the rest but do not suffer from smaller data sizes and are more resilient to drift in spectral transmittance. Even though we have shown in our previous work that re-calibrating the filters will overcome the drifts and wears in nanomaterials, but if the re-calibration is not a readily available option for the user, the SVM model offers acceptable accuracy and longer usability over time. On the other hand if speed is a consideration the ANN models would be the best choice, which turn out to perform well if enough data is provided. We also observed that ANN models with more number of layers seems to learn better from the available data. The choice of model depends on the application; for instance spectroscopy does not demand a fast real-time output but accurate and precise estimations. There are other applications especially in biology, for instance in DNA sequencing\cite{larkin2017length}, where the accuracy of the peak wavelength is not of importance as long as it is estimated close enough, but the time is of vital importance.

Furthermore, we have verified the possibility of modeling the drift of nanomaterials over time by observing the gradual changes in the filter functions, hence, being able to predict the filter function at later times, and thereby increase the accuracy of the ML algorithms and usability of the filters over longer periods of time. The efficacy of each ML model in our optical sensing problem reveals some key differences between this problem and other applications of ML in material science and engineering. The drift of nanomaterials properties for instance, which poses an important complication on the problem via evolving class distributions $i.e.$ gradually modifying the response function of the filters even though the classes $i.e.$ wavelengths remain the same. The other difference is in the feature selection. In optical sensing problem a very small number of features are chosen from optoelectronic properties (transmittance only in this case) of the nanomaterials, while in other areas the feature vector can be huge and very complex. The future work is to generalize the methods of this paper to broad-band optical spectra. All said, we believe that application of advanced data analytic algorithms has been very limited in optical sensing applications, and our findings can open up a new path for designing new generation optical detectors by harnessing advanced data analyzing algorithms/ ML techniques and significantly transform the field of high-accuracy sensing and detection using cyber-physical approaches.

\section{Computational Details}
\label{sec:methods}
\textbf{Data Structure.} The analysis of our data were performed on transmittance values measured over a wide spectral range, $351 nm<\lambda<1100 nm$) for each of the 11 nanomaterial filters, as well as 110 repetitions of these wavelength-dependent data. As mentioned in previous article, the repeated data was acquired to account for drifts, fluctuations, and other variations commonly observed in physical measurements especially in nanomaterial-based systems, which tend to be sensitive to their environments \cite{hejazi2019transition, hejazi2019wavelength}. On the other hand larger training data usually results in better performance of most ML algorithms. From the mentioned 110 spectra of each nanomaterial filter, 100 of them were labeled as "training data" or sample-label pairs and used for training the models ($M=750\times{100}=75000$ training samples). The other 10 spectra per filter were labeled as initial "test samples" ($M^{'}=750\times{10}=7500$ test samples or original test samples), and were used only for testing the "trained" models. In another words the test samples were not part of the training process and the machine learning models did not "see" these samples until the testing step. 

In our classification problem there are $N$ different classes: one per wavelength, and we are trying to classify our transmittance data into these $N$ classes. Here, we will concisely introduce each ML method and give their mathematical equations; also we will mention the number of parameters that are being trained in each model. Computations are carried out in Python 3.7 using a 2.5 GHz Quad-core Intel Core i7.

\vspace{0.1in}
\textbf{Bayesian Inference.} The filters are not chemically independent from each other, for they are mixtures from different proportions of the same two nanomaterials; so for computational purposes we assume independence between their outcomes, and model them with Naive Bayes algorithm \cite{sahami1996learning, hejazi2019transition}. The Bayesian inference for wavelength estimation problem can be formulated as follows: Let $\Lambda = \{\lambda_{1},..., \lambda_{i},...,\lambda_{N}\}$ be $N$ different wavelengths in desired spectral range and with specified granularity ($i.e.$ 351--1100 $nm$ with 1$nm$ step in this study), and $T=\{t_{1},...,t_{i},...,t_{Q}\}$ be the transmittance vector of $Q$ filter values (\emph{i.e} $Q=11$ when all of the filters are used in this study). Employing the Bayesian inference, the probability of the monochromatic light having the wavelength $\lambda_{j}$ based on the observed/recorded transmittance vector $T$ is called \emph{posterior} probability 

\begin{equation}
    P(\lambda_{j} \mid T)=\frac{P(T\mid \lambda_{j})P(\lambda_{j})}{P(T)},
\end{equation}


which is the conditional probability of having wavelength $\lambda_{j}$ given transmittance vector $T$; $P(\lambda_{j})=\frac{1}{N}$ is the \emph{prior} probability which is a uniform weight function here since all of the wavelengths are equally-likely to happen; $N$ is the total number of quantifiable wavelengths in the range under study. Moreover, $P(T\mid \lambda_{j})=\prod_{i=1}^{Q}{P(t_{i}\mid \lambda_{j})}$ is the probability of observing transmittance data $T$ given wavelength $\lambda_{j}$, and is called the \emph{likelihood}, which is the probability of having transmittance vector $T$ if wavelength is $\lambda_{j}$; ${P(T)}$ is the \emph{marginal} probability which is the same for all possible hypotheses that are being considered, so acts as a normalization factor to keep the posterior probability in the range of 0 to 1.

Individual $P(t_{i}\mid \lambda_{j})$ values are assumed to be Gaussian normal distributions for each filter at each wavelength, and their mean values and standard deviations were calculated from the training data ($i.e.$ the 100 measured transmittance spectra) collected for each filter at each wavelength. Finally, given the measured transmittance sample $T'$ (a vector of $Q=11$ elements -- one transmittance value per filter at an unknown wavelength), the wavelength $\lambda^*$ of the unknown monochromatic light is estimated by choosing the value of $\lambda_j$ that maximizes the posterior probability $P(\lambda_{j} \mid T')$:

\begin{equation}
   \lambda^* = \argmax_{\lambda_{j}} P(\lambda_{j} \mid T'),
\end{equation}

This optimization called the maximum \textit{a posteriori} (MAP) estimation \cite{bassett2016maximum, bernardo2001bayesian, lee1989bayesian}. To clarify the estimation steps further we notice the Bayesian inference finds probability of the combined Q measured test transmittance values named $T'$ in the entire wavelength spectrum. According to MAP estimation the wavelength at which this probability is maximum is indeed the estimated wavelength in Bayesian inference. Though, from machine learning point of view no parameters are being learned in Bayesian inference, but considering the parameters of Gaussian distribution that we calculate in this method we can say overall $2QN=16500$ parameters are being learned in this approach, $N$ mean values and $N$ standard deviations from the training data.

\vspace{0.1in}
\textbf{k-Nearest Neighbors.} There are two main categories for kNN: (1) centroid-based, which a new test sample is classified by the distance of its feature values with the average ($i.e.$ centroid) of features of all training samples that belong to the same each class, and (2) by-instance-based, which is the standard kNN approach, in which a new case is classified by a majority vote of its neighbors, with the case being assigned to a class that is most common among its $k$ nearest neighbors measured by a distance function. If $k$ is $1$, then the case is simply assigned to the class of its nearest neighbor. Since kNN model with small $k$ is prone to over-fitting, usually a finite odd number is chosen for k. There are various kinds of distance functions which from them the four famous distance functions: \emph{Euclidean}, \emph{Manhattan}, \emph{Chebyshev}, and \emph{Minkowski} are used in this study, but only the results of \emph{Euclidean} distance function is presented which is the classical presentation of distance and is given by $df(X,Y) = \sqrt{\sum{(x_i - y_i)^2}}$. Here, $X$ refers to each sample in the training set and $Y$ refers to the unknown (test) sample. To apply it to our data we need to find distance of a new transmittance vector of $Q=11$ elements, $T'=\{t'_{1},...,t'_{i},...,t'_{Q}\}$, with all known transmittance vectors $T=\{t_{1},...,t_{i},...,t_{Q}\}$ that are already known and labeled in the training set, so the distance function is

\begin{equation}
    df(T,T') = \sqrt{\sum_{i=1}^{Q}(t_i - t'_i)^2}.
\end{equation}

The distance between $T'$ and all $M$ training samples is calculated, and the M calculated distance values are sorted from smallest to largest using a typical sorting algorithm. Afterwards, the $k$ nearest neighbors $i.e.$ wavelengths that have smallest distance values from the test $T'$ are found, which are the arguments of the first $k$ numbers of the sorted list. Each nearest neighbor is assigned a uniform weight of $1/k$, and the $k$ neighbors are classified. Then, the test case $T'$ is assigned to the group with largest vote or population. In order to find the best $k$ for our system we tried different values for $k$ in the range $k=[1,20]$, and picked $k=7$ which performed the best. As mentioned before, kNN is a non-parametric classification algorithm so, no parameters are being learned in kNN.

\vspace{0.1in}
\textbf{Artificial Neural Networks.} In an ANN model each layer is made from a fixed number of neurons. The output of each neuron is linear combination of corresponding input followed by a non-linear activation function such as logistic \emph{sigmoid} or \emph{softmax}. These layers are connected by weight matrices, so that for an input sample $T$, by performing layer by layer matrix multiplication we would like to get as close as possible to the the real label ($y$ value) of that sample. Let's show a three layer ANN model (with 1 hidden layer) with the layers by $a^{(1)}$, $a^{(2)}$ and $a^{(3)}$. To calculate each layer $a^{(l)}$, $l>1)$, first, a matrix multiplication is performed between previous layer and the hypothesis matrix $\theta^{(l-1)}$ to get $z^{(l)}$; then, an activation function $g(z)$ (usually sigmoid function $g(z)=\frac{1}{1+e^{-z}}$) is applied on $z^{(l)}$ which results in $i^{th}$ layer. 


First we construct $H$ linear combinations of the input variables $T=\{t_{1},...,t_{i},...,t_{Q}\}$ in the form

\vspace{-0.4in}
\begin{align}
      & z_{j}^{(2)} = \sum_{i=1}^Q w_{ji}^{(1)}t_i + w_{j0}^{(1)} \\
      & a_{j}^{(2)} = g(z_{j}^{(2)}) 
\end{align}

where, $j=1,...H$, and $H$ is the size of first hidden layer; and the superscript (1), (2) indicate that the corresponding parameters are in the first or second layer of the network. $w_{ji}$ is corresponding weights. $w_{j0}^{(1)}$ is referred as biases; $z_j$ are called activations and $g(z)$ is the mentioned nonlinear activation function. At each layer of ANN, there is such a transformation; for example in three layer ANN which includes only 1 hidden layer, the elements of the third layer will take the form

\vspace{-0.4in}
\begin{align}
      & z_{n}^{(3)} = \sum_{j=1}^{H} w_{nj}^{(2)}a_{i}^{(2)} + w_{n0}^{(2)}\\
      & a_{n}^{(3)} = g(z_{n}^{(3)}) 
\end{align}

where, $n=1,..,N$ and $N$ is the total number of outputs. $a_{n}^{(3)}$ is the final output of the hypothesis that is going to be compared with the known target wavelengths The bias parameters can be absorbed into the set of weight parameters by defining an additional input variable $t_0$ whose value is kept fixed at $t_0=1$, and the same for other layers, so we combine these various stages to give the overall network function that, for sigmoidal output unit activation functions, takes the form

\begin{equation}
     y_{m}(T,W) = g \Big(\sum_{j=0}^{H1} w_{kj}^{(2)} g \Big(\sum_{i=0}^Q w_{ji}^{(1)}t_i \Big) \Big)
\end{equation}

Given a training set comprising a set of input vectors ${T_m}$, where $m=1,..,M$, together with a corresponding set of target vectors ${r_m}$, we minimize the error function (also called optimization objective or cost function) which for sigmoidal case using Lagrange multiplier method it transforms into 


\begin{equation}
    E(W)=-\sum_{m=1}^{M} \Big[r_{m}\ln y_{m} + (1-r_{m})\ln (1-y_{m}) \Big].
\end{equation}

or more explicitly

\begin{equation}
    E(W)=-\sum_{m=1}^{M} \sum_{n=1}^{N}\Big[r_{mn}\ln y_{mn} + (1-r_{mn})\ln (1-y_{mn}) \Big].
\end{equation}

where $y_{mn}$ denotes $y_{n}(X_m,W)$. So far, we have explained the FeedForward propagation. At first there is a large cost because the model is not trained yet. An important step in ANN learning process is called Backpropagation which unlike the FeedForward propagation explained above, it propagates from last layer and stops on second layer. 
In Backpropagation, each of the weight parameters are updated a small amount proportional to the gradient of cost (error) function with respect to that weight parameter. The proportion factor is called \emph{learning rate} that defines updating rate for each parameter in Backpropagatoin. The training process happens by iterating many cycles, that in each cycle, we perform the FeedForward propagation, calculate the gradient, update the parameters of $\theta$ matrices during Backpropagation, and repeat the loop until the model can classify the training data (in output layer) with desired level of accuracy. Afterwards, the trained model can be used to classify the incoming new sample via a few simple matrix multiplications. The total number of parameters in ANN is equal to elements of the weight matrices that each layer is multiplied into plus a single bias element at each hidden layer. In this study, we examined two architectures of ANN: a three layer network with 1 hidden layer $H1=100$ neurons between an input layer of $Q=11$ and output layer of $N=750$ neurons, and a four layer network with 2 hidden layers of sizes $H1=100$ and $H2=400$. In first case the number of parameters is $(Q+1)H1 + (H1+1)N = 76950$. In the second architecture the total number of parameters is $(Q+1)H1 + (H1+1)H2 + (H2+1)N = 342350$. In the next section we will estimate computational complexity of each machine learning technique which is an indicator of the testing time. In this project, the Python's PyTorch package is used for building the ANN model. The training step of our ANN models are carried out using Northeastern University's Discovery cluster.


\vspace{0.1in}
\textbf{Support Vector Machines. }We begin our discussion of SVMs by returning to the two-class classification problem using linear models of the form

\begin{equation}
    \label{eq:4}
    y(T) = w^{\intercal}\phi(T)+b
\end{equation}

where $\phi(T)$ denotes a fixed feature-space transformation, and we have made the bias parameter b explicit.\cite{bishop2006pattern} The training dataset comprises M input vectors $T_1, ..., T_M$ with corresponding target values $r_1, ..., r_M$ where $r_{m} \in \{-1,1\}$, and new data points $T'$ are classified according to the sign of $y(T')$. We shall assume for the moment that the training dataset is linearly separable in feature space, so that by definition there exists at least one choice of the parameters
W and b such that a function of the form Equation (\ref{eq:4}) satisfies $y(T_m) > 0$ for points having $r_m = +1$ and $y(T_m) < 0$ for points having $r_m = -1$, so that $r_{m}y(T_m) > 0$ for all training data points. \cite{bishop2006pattern} In support vector machines the decision boundary is chosen to be the one for which the margin is maximized by solving 

\begin{equation}
    \argmax_{w,b} \Big\{ \frac{1}{\mid w \mid}\min_{m}\big [r_m(w^{\intercal}\phi(T_m)+b)\big] \Big\}
\end{equation}

where we have taken the factor $\frac{1}{\mid w \mid}$ outside the optimization over $m$ because $W$ does not depend on $m$.\cite{bishop2006pattern} On the other hand there are different kernel tricks to create a non-linear models, hence create larger feature space by a non-linear kernel function $k(T_i,T_j)=\phi(T_i)^{\intercal}\phi(T_j)$. This allows the algorithm to fit the maximum-margin hyperplane in a transformed feature space by replacing the Equation (\ref{eq:4}) with

\begin{equation}
    \label{eq:6}
    y(T) = \sum_{m=1}^{M}\alpha_{m}r_{m}k(T,T_m)+b
\end{equation}

The transformation may be non-linear and the transformed space high-dimensional.\cite{boser1992training, aiserman1964theoretical, cortes1995support} The RBF kernel for example uses a Gaussian distribution for each feature $T_i$ and creates $M$ different features using kernel function $k(\vec{T_i},\vec{T_j})=e^{-\gamma \mid (\vec{T_i}-\vec{T_j} \mid ^2}$ for $\gamma > 0$. 

In this work, apart from linear SVM, we also tried different kernel functions as \emph{Polynomial}, \emph{Gaussian Radial Basis Function} (RBF), \emph{sigmoid}, and \emph{Hyperbolic Tangent}, but we only report the results of the linear and RBF models, since they performed slightly better that the other models, and their outputs are pretty much the same for our data; for that reason we are presenting only one set of results for SVM which is for linear SVM and SVM with RBF kernel. The number of parameters to be learned is $N(Q+1)=9000$ for linear model, and $N(M+1)\sim 56\times10^6$ for RBF model, where $N=750$ is the number of classes, $M=75000$ is number of training samples and $Q=11$ is dimension of each training sample. Even though these numbers seem pretty large specially for RBF kernel, but most of these parameters are zero, and the calculation is carried out using sparse matrix of parameters. In fact, SVM kernels are called sparse kernel machines. \cite{bishop2006pattern} In this project, the Python's SciKit package is used for building the SVM model.

\vspace{0.1in}
\textbf{Time Complexity Analysis. }As given above, we have $N=750$ classes of all possible wavelengths, $Q=11$ filters as feature number and totally $M=75000$ samples for training. Once trained, we care more about their inference efficiency. The time complexity is analyzed as following. 

For Bayesian estimation, for each data point, the conditional distribution $P(T|\lambda_j)$ is calculated with all possible wavelengths which is $N$. The production for joint probability takes N operation as well. However power operation is included in Gaussian distribution density function. As we compute this density across whole spectrum, exponent in this operation will be $N$ related; therefore, each iteration of the implementation takes $O(N)$. So totally, Bayesian takes $O(QN)$ time.

For KNN, calculating the distance with all training data takes $O(QM)$ time. After that, finding the $k=7$ minimum values and their indices takes $O(kM)$ time, so the overall complexity is in the order of $O(kM)$.

For ANN, the computation cost from layer $i$ to layer j is $H_iH_j$. For multiple layer version, the time cost can be generalized as $O(QH_1+ \sum_{i=1}^{L} H_iH_{i+1})$, where $H_i$ stands for the hidden neuron numbers at layer $i$, L stands for the total layer numbers (except input). 
As the $Q$ and $H$ will be data size-independent, this method is supposed to be much faster than the other methods. In our ANN architecture the number of neurons increases almost by order of magnitude as we go from input to output layer, so the complexity is dominated by the last layer. $N$ is output layer size and if we denote the last hidden layer size by $H_L$, the time complexity will be in the order of $O(H_LN)$.

For SVM, for each query the kernel operation is across all support vectors within training data. 
The inference complexity for linear and RBF model will be $O(QM_{sv})$ since we are solving the dual form here; $M_{sv}$ stands for the number of support vectors, which most of the times will be much less than $M$ but it can also be upto $M$, so we can show its upper limit as $O(QM_{sv})$.

\begin{table}[h]
    \centering
    \begin{tabular}{c|c c c c}
                        ML Model & Bayesian & kNN        & SVM & ANN \\
                        \hline
        Time Complexity & $O(QN)$  & $O(kM)$    & $O(QM)$ & $O(H_LN)$
    \end{tabular}
    \caption{Time Complexity of ML method}
    \label{tab:TMcomplexity_table}
\end{table}

The theoretical complexity estimations are in agreement with the measured time required for testing each sample (see \figref{AbserrorTesttime}(b).

\section{Acknowledgment}
DH and SK acknowledges financial support from NSF ECCS 1351424, and a Northeastern University Provost's Tier 1 Interdisciplinary seed grant.


\section{Author Information}
\label{sec:authorinfo}
\textbf{ORCID\\
Davoud Hejazi:0000-0002-5215-6395\\
Shuangjun~Liu: 0000-0002-2717-5789\\
Amirreza Farnoosh: 0000-0002-3766-2310\\
Sarah Ostadabbas: 0000-0002-2216-9988\\
Swastik Kar: 0000-0001-6478-7082}


\bibliographystyle{achemso}
\bibliography{paper}


\end{document}